\documentstyle[preprint,aps,epsfig,floats]{revtex}
\draft
\begin{document}
\date{\today}
\preprint{\vbox{\hbox{UA/NPPS-14-01}\hbox{RUB-TPII-10/01}}}
\title{RESUMMATION OF PERTURBATIVELY ENHANCED\\
GLUON RADIATIVE CORRECTIONS \\}
\author{{\bf A. I. Karanikas}${}^{1}$\thanks{E-mail:
        akaranik@cc.uoa.gr},
        {\bf C. N. Ktorides}${}^{1}$\thanks{E-mail:
        cktorid@cc.uoa.gr},
        {\bf N. G. Stefanis}${}^{2}$\thanks{E-mail:
        stefanis@tp2.ruhr-uni-bochum.de}
        }
\address{${}^{1}$ University of Athens,
                  Department of Physics,
                  Nuclear and Particle Physics Section,   \\
                  Panepistimiopolis,
                  GR-15771 Athens, Greece                 \\
                  [0.1cm]
         ${}^{2}$ Institut f\"ur Theoretische Physik II,
                  Ruhr-Universit\"at Bochum,
                  D-44780 Bochum, Germany                 \\
         }
\maketitle
\begin{abstract}
By remaining strictly within the confines of QCD, i.e.,
{\it without} invoking the parton model or any other exogenous
element, we identify and resum perturbatively to leading order,
along with a correction term, enhanced radiative gluon contributions
to the Drell-Yan type ($q\bar{q}$ pair annihilation) and
deep-inelastic-scattering type ($eq\rightarrow eq+X$)
cross-sections. The key feature of the adopted approach is the
recasting of QCD in terms of a space-time mode of description,
which employs a path-integral formulation of field theories,
as originally implied in works of Fock, Feynman and Schwinger.
\end{abstract}
\newpage

\section{Introduction}

Studies surrounding the infrared (IR) behavior of transition
amplitudes in gauge field theories present special interest not
only for the fundamental -- of interpretational nature --
questions they pose but also for the calculable physical
implications they entail. To become more specific and, at the same
time, identify the object of interest in the present investigation,
we are here alluding to resummation procedures of perturbatively
enhanced pure QCD contributions to physical cross-sections.
These matters have come under systematic scrutiny by Sterman and
collaborators \cite{Ster87,CS93,CLS97}, following earlier realizations
\cite{AV79} on the physical relevance of soft gluon radiation.

The wide acceptance of QCD as the microscopic theory of the strong
interactions, despite the fact that its fundamental field quanta do
not register as observed asymptotic states, rests largely on one's
ability to isolate and identify its contribution to a given hadronic
process by means of factorization theorems. With respect to the impact
of IR properties of QCD on measurable quantities, the studies
conducted in \cite{Ster87,CS93,CLS97}, which commence at the hadronic
level and methodically proceed to the microscopic one \cite{CSS89},
encourage the idea to pursue IR issues through a direct reference to
QCD, i.e., through considerations which base themselves on the Green's
functions of the theory. Recognizing, at the same time, the key role
played by the parton model in connection with applications of QCD, we
shall employ for its description a Fock-Feynman-Schwinger
\cite{Foc37,Fey50,Sch51}, or equivalently worldline casting
\cite{KK92,Str92,SS94} (for recent reviews, we refer to
\cite{Schu01,SiTj02}), wherein a space-time interpretation --
as opposed to a Feynman diagrammatic one -- of fundamental matter
field quanta propagation is attained \cite{KK95}. In this way, one
aspires to achieve consistency with the parton model mode of
description without, however, invoking its probabilistic content.
Our efforts, in this paper, will be directed towards determining pure
QCD contributions related to processes induced by electroweak
interactions -- Drell-Yan (DY) and deep inelasting
scattering (DIS), in particular.

For the IR domain of QCD, a central issue of concern is where the
perturbative picture ends and to what extent non-perturbative effects
start to become dominant (for a review and references, we refer to
\cite{Ste99}). As a matter of fact, the very concept of soft gluon
radiation looses its meaning below some momentum scale of the order
of $\Lambda_{\rm QCD}$ (recent discussions of these issues are given
in \cite{SSK00,KS01}). One consistent approach for attacking this
problem strictly within the framework of the microscopic theory
itself has been articulated by Ciafaloni \cite{Cia89}. As already
mentioned, our own considerations will rely upon the worldline
casting of QCD, keeping a more pragmatic course and focusing
our attention on cross-section expressions.

One might reasonably wonder as what constitutes the innovative
tool, provided by the worldline casting of QCD, that facilitates
calculations which ``flirt'' with the edge of the perturbative
domain of the theory. A direct answer {\it can} be given at this point
though its full content will become more transparent through the main
exposition in the sections to follow. In short, it is the ability
to isolate a special set of space-time paths having a very simple
geometrical profile shared in a restricted (but directly relevant to
the physics of the process) neighborhood by each and every contour
entering the path integral. The single (multiplicative) renormalization
constant carried by this special family of paths automatically
factorizes their contribution to amplitudes/cross-sections given that
it also accompanies the rest of the paths. The more complex geometrical
structure of the latter simply implicates additional ultraviolet (UV)
singularities which can be absorbed into conventional wave- function
and coupling-constant renormalizations. This clean, geometrically
based, argument, which singularly underlines the worldline description,
not only constitutes a notable simplification over the type of
reasoning that has been promoted by Collins et al. \cite{CSS89} within
the Feynman diagram perturbative approach to realize factorization in
QCD, but, potentially, it paves the way to further applications.
We especially have in mind issues pertaining to projections of our
approach into the non-perturbative domain of the theory.

Letting these comments suffice for an introductory exposition, we
now proceed to display the organization of this paper, which is as
follows.
In the next section, we exhibit the worldline expression for the full
fermionic Green's function and subsequently employ it to construct
corresponding expressions both for DY and DIS type QCD
amplitudes/cross-sections. Section 3 furnishes, with the aid of an
Appendix, our basic calculations associated with one virtual
gluon exchanges for a special set of trajectories. The resulting
expression explicitly reveals the threshold enhancement factor while
the task of virtual gluon re-summation is performed, via the aid of
the renormalization group, in Section 4.
Section 5 deals with the re-summation of contributions from
real gluons, whereas our conclusions are presented in Section 6.

\section{Basic worldline expressions for amplitudes and cross-sections}

Consider the full two-point (fermionic) Green's function in the
presence of an external gluonic field. The expression, in Euclidean
space-time,
\begin{eqnarray}
  iG_{ij}(x,y|A)
&=&
  \int_{0}^{\infty} dT {\rm e}^{-Tm^{2}}
  \int_{\stackrel{x(0)=x}{x(T)=y}} {\cal D}x(t)
  \left[ m-{1\over 2}\gamma\cdot \dot{x}(T) \right]
  P\exp\left( {i\over 4}\int_{0}^{T} dt \sigma_{\mu\nu}\omega_{\mu\nu}
  \right)
\nonumber\\
&& \times
   \exp\left[ -{1\over 4} \int_{0}^{T} dt\dot{x}^{2}(t) \right ]
  P\exp\left [ ig\int_{0}^{T} dt\dot{x}\cdot A(x(t)) \right ]_{ij},
\label{eq:green}
\end{eqnarray}
displays the basic worldline features pertaining, more generally,
to $n$-point Green's functions\footnote{In this paper we shall be
exclusively concerned with electroweak vertex functions of quarks
which include gluon radiative effects.} and, by extension, to
amplitudes. Here, and below, $P$ denotes the usual path ordering
of the integrals. The first thing to point out is that a given path of
the matter field quantum, starting at $x$ and ending at $y$ between
respective ``proper-time'' values $0$ and $T$, also enters a Wilson
line factor. The latter, being the sole carrier of the dynamics,
separates itself from the rest of the factors in the path integral
which are associated with  geometrical properties of paths traversed
by spin-1/2 particle entities. The most notable such quantity is the
so-called spin factor
\cite{Pol90},
$
 P\exp \left[ (i/4)\int_{0}^{T} dt\,\sigma\cdot\omega
       \right ]
$,
where
$
 \omega_{\mu\nu}
=
 (T/2)(\ddot{x}_\mu \dot{x}_\nu-\dot{x}_\mu \ddot{x}_\nu)
$,
accounting, in a geometrical way, for the spin-1/2 nature of the
propagating particle.
Accordingly, our perturbative expansions should be perceived of
in terms of (Euclidean) space-time paths involving a ``proper time''
parameter and {\it not} in terms of Feynman diagrams.
As it turns out \cite{KK99}, in the perturbative context, the
structure of matter particle contours, entering the path
integral, is determined by the points, where a momentum
change takes place, i.e., points where a gauge field line
(real or virtual) attaches itself on the (fermionic) matter field
path.
The almost everywhere non-differentiability of these contours,
is residing precisely at these points.
A major effort, in this paper, will be devoted to the
extension of the worldline formalism to expressions for cross-sections
corresponding to the particular processes of
$q\bar{q}$
annihilation and
$e+q\rightarrow e+q+X$.

From the worldline point of view, the processes we intend to study
involve fermionic matter particle (quarks) paths that commence at $x$
and end at $y$, being forced to pass through an intermediate point $z$,
where a momentum transfer $Q$ takes place. This means that the Green's
(vertex-type) function we shall be dealing with has the following
form ($\Gamma_\mu$ denotes some Clifford-Dirac algebra element)
\begin{eqnarray}
  V_{\mu,ij}(y,z,x|A)
&=&
  G_{ik}(y,z|A) \Gamma_{\mu} G_{kj}(z,x|A)
\nonumber\\
&\!=\!&
  \int_{0}^{\infty} dT {\rm e}^{-Tm^{2}}\!\int_{0}^{T} ds \!
  \int_{\stackrel{x(0)=x}{x(T)=y}} {\cal D}x(t)
  \delta \left( x(s)-z \right)
  \Gamma_{\mu}\left( \dot{x},s \right)
  \exp\left[ -{1\over 4}\int_{0}^{T} dt\,\dot{x}^{2}(t) \right]
\nonumber\\
&&\times
  P\exp\left[ ig\int_{0}^{T}dt\, \dot{x}(t)\cdot A(x(t)) \right]_{ij},
\end{eqnarray}
\label{eq:vertex}
where
\begin{eqnarray}
  \Gamma_{\mu}\left( \dot{x},s \right)
& \equiv &
\left[ m-{1\over 2}\gamma\cdot \dot{x}(T) \right]
  P\exp\left( {i\over 4}\int_{s}^{T} dt\, \sigma\cdot\omega \right)
  P\exp\left( {i\over 4}\int_{0}^{s} dt\, \sigma\cdot\omega\right )
\nonumber\\
&& \times
  \Gamma_{\mu} \left [ m-{1\over 2}\gamma\cdot \dot{x}(s) \right].
\end{eqnarray}
\label{eq:gamma}
It is especially important to realize that in our approach off-shellness
is naturally parameterized in terms of the finite size of the matter
particle contours and realistically accounts for the fact that quarks
reside inside a hadron ($m$ can be viewed as an effective quark mass).

Going over to momentum space, we write
\begin{eqnarray}
  \tilde{V}_{\mu,ij}(p,p'|z|A)
= &&
  \int_{0}^{\infty} dT {\rm e}^{-Tm^{2}} \int_{0}^{T} ds
  \int_{}^{}{\cal D}x(t)\delta \left( x(s)-z \right)
  \Gamma_{\mu} \left( \dot{x},s \right)
\nonumber\\
&& \!\!\!\!\!\times \;
   \exp\left[ -{1\over 4}\int_{0}^{T} dt\, \dot{x}^{2}(t)
  +ip\cdot x(0)+ip^{\prime}\cdot x(T) \right]
\nonumber\\
&& \!\!\!\!\!\times \;
  P\exp\left[ ig\int_{0}^{T} dt\, \dot{x}(t)\cdot A(x(t)) \right]_{ij}
\nonumber\\
\equiv &&
  \sum_{C^{z}}\tilde\Gamma[C^{z}]P
  \exp \left[  ig\int_{C^{z}}^{} dx\cdot A(x) \right]_{ij},
\label{eq:vertex_mom}
\end{eqnarray}
where $C^{z}$ denotes a generic path forced to pass through point $z$,
at which the momentum $Q$ is imparted.

For a process of the type (DY)
$q+\bar{q}\rightarrow$
lepton pair + X the ``amplitude'' expression reads
\begin{eqnarray}
  \Delta^{(\text{DY})}_{\mu,ij}
& = &
  \bar{v}(p^{\prime},s^{\prime})(-i\gamma\cdot p^{\prime}+m)
  \tilde{V}_{\mu,ij}(i\gamma\cdot p+m)u(p,s)
\nonumber\\
& \equiv &
  \sum_{C^{z}}\tilde{I}^{(\text{DY})}_{\mu,p^{\prime}p}[C^{z}]
  P\exp\left[ ig\int_{C^{z}} dx\cdot A(x) \right]_{ij}
\label{eq:DY_amplitude}
\end{eqnarray}
with the second, comprehensive, expression to be understood
having recourse to Eq.~(\ref{eq:vertex_mom}).

Similarly, for the DIS-type process
$e+q\rightarrow e+q$ + X, one writes
\begin{eqnarray}
  \Delta^{(\text{DIS})}_{\mu,ij}
& = &
  \bar{u}(p^{\prime},s^{\prime})(i\gamma\cdot p^{\prime}+m)
  \tilde{V}_{\mu,ij}(i\gamma\cdot p+m)u(p,s)
\nonumber\\
& \equiv &
  \sum_{C^{z}}\tilde{I}^{(\text{DIS})}_{\mu,p^{\prime}p}[C^{z}]
  P\exp\left[ ig\int_{C^{z}}dx\cdot A(x) \right]_{ij},
\label{eq:DIS_amplitude}
\end{eqnarray}
with the replacement $p^{\prime}\rightarrow -p^{\prime}$
to be also made in Eq.~(\ref{eq:vertex_mom}).

\begin{figure}
\centering\epsfig{file=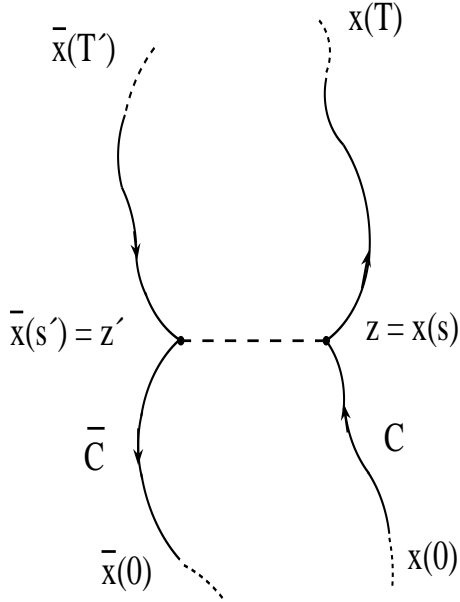,height=8cm,width=6.0cm,silent=}
\vspace{0.5cm}
\caption{\footnotesize
                Illustration of two conjugate contours $C$ and $\bar{C}$
                entering the worldline path integral, ``talking'' to each
                other at points $z$ and $z^{\prime}$, where the momentum
                transfer for the physical process takes place. The distance
                $|z-z^{\prime}|$ is referred to as the impact parameter.}
\label{fig:paths1}
\end{figure}

For the cross-section, we need to employ the following quantity,
which we implicitly display in Minkowski space-time after
straightforward adjustments,
\begin{eqnarray}
  \Delta^{(\alpha)\dagger}_\mu \Delta^{(\alpha)}_\nu
& = &
  \sum_{\bar{C}^{z^{\prime}}}
\everymath{\displaystyle}
  \sum_{\rule{0in}{1.4ex}{C^{z}}}
  \tilde{I}^{(\alpha)\dagger}_{\mu,p^{\prime}p}[\bar{C}^{z^{\prime}}]
  \tilde{I}^{(\alpha)}_{\nu,p^{\prime}p}[C^{z}]
\nonumber\\
&& \times
  Tr \left\{
  \bar{P}
  \exp \left[ ig\int_ { \bar{C}^{z^{\prime}} }
  dx^{\prime \mu} A_{\mu}(x^{\prime}) \right]
  P
  \exp \left[ -ig\int_{C^{z}} dx^{\mu} A_{\mu}(x) \right]
     \right\},
\label{eq:crossection}
\end{eqnarray}
where $\bar{P}$ denotes anti- path ordering and the index $\alpha$
stands for either DY or DIS.
Even though not explicitly displayed, the cross-section acquires a
path-integral form, which has the following characteristics:
1) Paths $C^{z}$ and $\bar{C}^{z^{\prime}}$ are forced to pass through
points $z$ and $z^{\prime}$, respectively, where the momentum transfer
occurs (see Fig.~\ref{fig:paths1}).
The distance $b\equiv |z-z^{\prime}|$ serves as a measure of how far
apart the two conjugate contours can venture away from each other and
will be referred to as the impact parameter.
2) The traversal of $\bar{C}^{z^{\prime}}$ is made in the opposite
sense relative to $C^{z}$, pretending the two paths join at one
end (using translational invariance), while allowing the other two
ends of the contour to close at infinity, giving rise to the formation
of a Wilson loop.
3) Under the circumstance just described, the Wilson loop formation
guarantees the gauge invariance of the expression for the cross
section.
On the other hand, by keeping the contour lengths finite, but very
large, thereby placing the quarks off-mass-shell, gauge invariance
will still continue to hold to the order of approximation we
employ in our computations, given that the off-mass-shellness also
serves as an IR cutoff.

%

\begin{figure}
\centering \epsfig{file=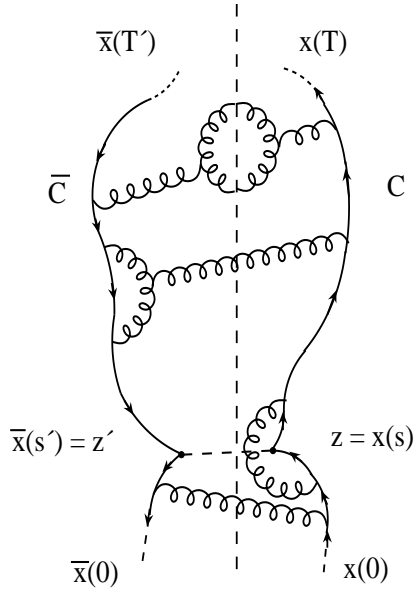,height=8cm,width=6.0cm,silent=}
\vspace{0.5cm}
\caption{\footnotesize
                Virtual gluon radiative corrections of various
                sorts and ``real'' gluon lines with their ends
                attached on each of the two depicted contours at
                the cross-section level.}
\label{fig:paths2}
\end{figure}

Up to this point our considerations have been centered around the
geometrical profile of the worldline casting of QCD.
In terms of our comprehensive formulas (\ref{eq:DIS_amplitude}) and
\ref{eq:crossection}) for the amplitude and cross-section,
respectively, the primary emphasis has been placed on the first
factors of the respective sums, the main conclusion being that the
relevant contours entering the path integral are {\it open} for the
amplitude and {\it closed} (or almost so) for the cross-section.
Armed with this information, we now turn our attention to the Wilson
factor which contains all the dynamics of the given process.
The obvious task in front of us is to assess its implications once the
gauge fields are quantized, i.e., once the Wilson factor is inserted
into a {\it functional} integral weighted by the exponential of the
Yang-Mills action.
We display the quantity of interest as follows
\begin{eqnarray}
  {\cal W}
& = &
  \left\langle
              Tr \left\{ \bar{P} \exp\left[ ig\int_{\bar{C}^{z^{\prime}}}
              dx^{\prime \mu} A_{\mu}(x^{\prime}) \right]
                 \right\}_{\rm A}
  \left\{P \exp \left[ -ig\int_{C^{z}}dx^{\mu} A_{\mu}(x)
                \right]
  \right\}_{\rm A}
  \right\rangle
\nonumber  \\
& \equiv &
  \langle Tr(U^{\dagger}(\bar{C}^{z^{\prime}})U(C^{z}))\rangle.
\label{eq:wilson}
\end{eqnarray}
In the above expression, $\{\cdot\cdot\cdot\}_{\rm A}$ signifies
the expectation value with respect to the gauge field functional
integral which, in this work, will be considered in the context of
perturbation theory.
Note in the same context that a virtual gluon attaching itself with
both ends to the fermionic worldline, entering the amplitude,
corresponds to a correlator between a pair of gauge fields originating
from the expansion of the Wilson factor.
On the other hand, for an emitted ``real'' gluon from the fermionic
line, the correlator is between an ``external'' and a Wilson-line
gauge field.\footnote{One will, of course, also encounter correlators
that involve gauge fields from the non-linear terms of the Yang-Mills
action. These, however, do not enter the leading logarithmic
considerations.}
The overall situation is depicted in Fig.~\ref{fig:paths2}.
At the cross-section level, now, ``real'' gluons are integrated with
respect to ``propagators'' linking together the two conjugate
contours, while their polarization vectors are summed over
(cut propagators).
This is precisely what $\langle\cdot\cdot\cdot\rangle$ signifies
in the last equation, as it covers both Wilson line factors.

In the light of the above remarks, let us proceed to display
the first-order (in perturbation theory) expression for ${\cal W}$,
which receives contributions both from virtual gluons, viz.; ones
attached at both ends either to worldline contour $C^z$ or to contour
$\bar{C}^{z^{\prime}}$, as well as from ``real'' gluons linking these
contours to each other (cf. Fig.~\ref{fig:paths2}).
This expression reads
\begin{eqnarray}
  {\cal W}^{(2)}
=
  {\rm Tr}\,I
& - &
     g^{2}C_{\rm F}\int_{0}^{T} dt_{1}
                   \int_{0}^{T} dt_{2}\,
                   \theta \left( t_{2}-t_{1} \right)
                   \dot{x}^{\mu} \left( t_{2} \right)\,
                   \dot{x}^{\nu} \left( t_{1} \right)\,
                   D_{\mu\nu} \left( x(t_{2})-x(t_{1}) \right)
\nonumber\\
&-&
     g^{2}C_{\rm F}\int_{0}^{T^{\prime}}dt^{\prime}_{1}
               \int_{0}^{T^{\prime}}dt^{\prime}_{2}\,
               \theta \left( t^{\prime}_{1}-t^{\prime}_{2} \right)
               \dot{x}^{\prime \mu} \left( t^{\prime}_{2} \right)\,
               \dot{x}^{\prime \nu} \left( t^{\prime}_{1} \right)\,
               \bar{D}_{\mu\nu} \left( x(t^{\prime}_{2})-x(t^{\prime}_{1}) \right)
\nonumber\\
&-&
     g^{2}C_{\rm F}\int_{0}^{T}dt
                   \int_{0}^{T^{\prime}}dt^{\prime}\,
                   \dot{x}(t)\cdot\dot{x}^{\prime} \left( t^{\prime} \right)\,
                   D_{\rm cut} \left( x(t)-x(t^{\prime}) \right)
                 + {\cal O}\left(g^{4} \right).
\label{eq:1orderwilson}
\end{eqnarray}
It becomes obvious from their structure that the first two non-trivial
terms correspond to virtual gluon contributions  -- one per
conjugate branch --, while the third one is associated with
``real'' gluon emission.
Finally, concerning the gluon propagators entering the above
equation, we shall be employing their Feynman-gauge form
without loss of generality due to gauge invariance.
In particular we have, in $D$-dimensions,
\begin{equation}
  D_{\mu\nu}(x)
=
  -ig_{\mu\nu}\mu^{4-D}\int\frac{d^{D}k}{(2\pi)^{D}}
  \frac{{\rm e}^{-ik\cdot x}}
  {k^{2}+i0_+}
=
  g_{\mu\nu}{1\over 4\pi^{2}}\left( -\pi\mu^2 \right)^{(4-D)/2}
  \frac{\Gamma(D/2-1)}{(x^{2}-i0_{+})^{(D/2)-1}},
\end{equation}
\label{eq:virtualgluonpropagator}
whereas
\begin{equation}
  D_{\rm cut}(x)
=
  \mu^{4-D}\int\frac{d^{D}q}{(2\pi)^{D}}2\pi\delta(q^{2})\theta(q^{0})
  {\rm e}^{-iq\cdot x}
=
  {1\over 4\pi^{2}}\left( -\pi\mu^{2} \right)^{(4-D)/2}
  \frac{\Gamma(D/2-1)}{\left[ (x_{0}^{2}-i0_{+})^{2}-{\bf x}^{2}
                       \right]^{(D/2)-1}}.
\label{eq:realgluonemission}
\end{equation}

As already established by other methods, the perturbative
expansion (\ref{eq:1orderwilson}) is plagued by large threshold
logarithms leading to the need for factorization and resummation.
This is precisely the task we are about to undertake within the
framework we have developed and which {\it does not} take us
outside QCD.

%

\section{First-order virtual gluon corrections universal to the
vicinities of points \lowercase{$\bbox{z}$} and
\lowercase{$\bbox{z^{\prime}}$}}

The space-time mode of description of our worldline casting puts us
into the position to promote the following argument: The point
$z$ (or $z^{\prime}$), where the momentum transfer $Q$ is imparted,
marks the presence of a neighborhood around it, no matter how
infinitesimal in size this might be, whose geometrical structure is
shared by {\it all} fermionic paths entering the path-integral.
Specifically, there will be a derailment (cusp formation), whose
opening angle will be fixed unambiguously, since it is determined by
the momentum transfer. It follows that the contributions to the
amplitude and cross-section from the immediate vicinities of these
points is a common feature of all contours and eventually factorizes.
In this section we shall determine the first-order perturbative term
corresponding exactly to this factor.

Consider now the neighborhood of point $z$ on the contour $C^{z}$.
Expanding around this point, we write
\begin{equation}
  x^{\mu}(t)
=
  x^{\mu}(s) + t \dot{x}^{\mu}(s\pm 0) + \ldots
\label{eq:expoint}
\end{equation}
with
$v^{\mu}=\dot{x}^{\mu}(s-0)$ and $v^{\prime \mu}=\dot{x}^{\mu}(s+0)$
being entrance and exit four-velocities, respectively, with respect to
$z$.

Adjusting our notation by re-parameterizing the contour so that
the zero value is assigned to point $z$, the relevant
quantity to compute, to first perturbative order, becomes
\begin{eqnarray}
  U_{C,S}^{(2)}
=
  1
&-&
  g^{2}C_{\rm F}
  \left[   \int_{-\sigma}^{0} dt_{1}\int_{-\sigma}^{0} dt_{2}
           \theta \left( t_{2}-t_{1} \right)
           v^{\mu} v^{\nu} D_{\mu\nu} \left( vt_{2}-vt_{1} \right)
         + \int_{0}^{\sigma} dt_{1} \int_{0}^{\sigma} dt_{2}
           \theta \left( t_{2}-t_{1} \right)
\right.
\nonumber\\
&+&
\left.            v^{\prime\mu} v^{\prime\nu}
            D_{\mu\nu} \left( v^{\prime}t_{2}-v^{\prime}t_{1} \right)
          + \int_{0}^{-\sigma} dt_{1}\int_{0}^{\sigma}dt_{2}
            \theta \left( t_{2}-t_{1} \right) v^{\prime\mu} v^{\nu}
            D_{\mu\nu}\left( v^{\prime}t_{2}-vt_{1} \right)
  \right].
\label{eq:U2order}
\end{eqnarray}
It is clear that the above expression corresponds to the first term in
Eq.~(\ref{eq:1orderwilson}), which monitors virtual gluon
exchanges occurring on contour $C$.

From the above considerations it follows that the main contribution
to each double integral comes from the common limit
$t_{1},\,t_{2}\rightarrow 0$.
Suppose now, the other limit is to be determined by demanding that
its contribution to the integrals is of vanishing importance.
Then, such a requirement automatically isolates those contours,
whose only significant geometrical characteristic is that the
four-velocities to approach and depart from point $z$ are
fixed, denoted by $v^{\mu}$ and $v^{\prime\mu}$ respectively;
the same, of course, happens for point $z^{\prime}$, but in
reverse order.
This justifies the subscript $S$ in $U_{C,S}$, which stands for
``smooth''.
More complex geometrical properties such as non-differentiability
almost everywhere (which most certainly characterizes the vast
majority of paths) do not enter Eq.~(\ref{eq:U2order}).
The point is, on the other hand, that {\it every path} will share
the geometrical structure we are focusing on in some neighborhood of
the point $z$, or $z^{\prime}$, no matter how close to these points
one has to come.
At the same time, the UV singularities of this restricted set of paths
will entail expressions that solely depend on the two four-velocities
and the opening angle.
Paths of more complex geometrical structure, on the other hand, will
certainly exhibit these UV singularities plus additional
ones.\footnote{Actually, the standard UV singularities of perturbative
field theories associated with $\beta$-functions, coupling constant
and wave-function renormalization pertain to almost everywhere
non-differentiable paths.}
It naturally follows that our restricted set of trajectories, by
exclusively determining the corresponding (multiplicative)
renormalization constant will {\it factorize} from the rest of the
expression for the amplitude and/or the cross-section.
From dimensional considerations it follows that the omitted terms in
Eq.~(\ref{eq:expoint}) will contain negative powers of $\sigma$,
whose dimension in the denominator is ${\rm (mass)}^{2}$.
Neglecting their presence means that $\sigma$ should be very large in
magnitude and hence should be related to the IR cutoff, i.e.,
$\sigma\simeq\lambda^{-2},\,\lambda > \Lambda_{\rm QCD}$.

Going over to Minkowski space-time, there are two distinct
possibilities for defining an infinitesimally small neighborhood
around $z$.
The first one, to be labelled $(a)$, reads
\begin{equation}
  (x-x^{\prime})^{2}
=
  {\cal O}(\epsilon^{2}),\,\quad {\rm with} \quad
  v_\mu\simeq v^{\prime}_{\mu},\quad {\rm for}\;\,{\rm all}\quad \mu,
\label{eq:neighbor1}
\end{equation}
where $\epsilon(\leq Q^{-1})$ is a small length scale.
The second alternative, to be labelled (b), can be typically
represented by
\begin{eqnarray}
  (x-x^{\prime})^{2}
&\!\!=\!\!&
  {\cal O}(\epsilon^{2})\,{\rm with}\,
  |v-v^{\prime}|^{2}
=
  {\cal O}(\lambda^{2})\,
  {\rm but}\,(v_{+} - v^{\prime}_{+})
\simeq
  {\cal O}(Q)\,{\rm and}\,(v_{-} - v^{\prime}_{-})
\simeq
  {\cal O}\left( \frac{\lambda^{2}}{Q} \right)
\nonumber\\
&& \quad
  \!\!\!\!\!\!\!\!\!\!\Rightarrow (v_{+} - v^{\prime}_{+})(v_{-} - v^{\prime}_{-})
=
  {\cal O}(\lambda^{2})
\label{eq:neighbor2}
\end{eqnarray}
that is equivalently effected via the condition
$v_{+} \gg v^{\prime}_{+},\, v_{-} \simeq v^{\prime}_{-}$.
All in all, there are four different configurations:
$+\leftrightarrow -$ and prime$\leftrightarrow$no-prime
entering this case.

We characterize case $(a)$ as ``uniformly soft'', given that
the considered gluon exchanges take place in a neighborhood whose
smallness pertains to all directions. Case $(b)$, on the other hand,
will be referred to as ``jet'' since gluon emission occurs under
circumstances, where entrance and exit four-velocities differ between
them significantly along one or the other of the light-cone
directions. We point out that what here is termed as ``jet'' does not
coincide with the notion of Collins et al.~\cite{CSS89}.
The actual connection will become evident later on.

Let us commence our calculations by taking up the first ${\cal O}(g^{2})$
term entering the right hand side of Eq.~(\ref{eq:U2order}).
Since this only involves the branch of the contour $C^{z}$ entering
point $z$, we obtain the same expression regardless of whether or not a
uniformly soft or a jet configuration is being considered.
It reads
\begin{eqnarray}
  I_{1}
& = &
   \int_{-\sigma}^{0} dt_{1} \int_{-\sigma}^{0} dt_{2}\,
   \theta\left( t_{2}-t_{1} \right)v^{\mu} v^{\nu}
  D_{\mu\nu}\left( vt_{2}-vt_{1} \right)
\nonumber\\
&=&
  -\frac{1}{8\pi^{2}}\left( -\pi\mu^{2}L_{1}^{2} \right)^{(4-D)/2}\,
   \Gamma\left( \frac{D}{2}-1 \right)
   \frac{1}{D-3}\,\frac{1}{2-D/2},
\label{eq:I1}
\end{eqnarray}
where\footnote{Note that $v$ has dimensions of mass as our ``time'' parameter
$\sigma$ has dimensions of (mass)$^{-2}$.}
$L_{1}= \sigma|v|$.
The second term has the same structure (it involves the exiting branch of
$C^{z}$) and therefore produces a similar result:
\begin{equation}
  I_{2}
=
  -\frac{1}{8\pi^{2}}\left( -\pi\mu^{2}L_{2}^{2} \right)^{(4-D)/2}\,
   \Gamma\left( \frac{D}{2}-1 \right)
   \frac{1}{D-3}\,\frac{1}{2-D/2}
\label{eq:I2}
\end{equation}
with $L_{2}=\sigma |v^{\prime}|$.

A couple of remarks are in order at this point. First, even though the
length scales $L_1$ and $L_2$ are both large, being proportional to
$\sigma$, they will be of the same order of magnitude for case (a),
whereas for case (b), one scale will be negligible in comparison with
the other.
Accordingly, the total expression for the uniformly soft amplitude will
be twice as large as that of the jet-like one. This being said, we shall
denote the dominant length scale by $L(\simeq L_{1}$ and/or $L_{2})$,
when it enters our final expressions, and set it equal to
${1\over \lambda}$, recognizing that it is of the same order as the IR
cutoff. Second, in order to avoid the double counting resulting from the
fact that each branch has been ``cut-off'' at corresponding distances
away from $z$ (implying that gluon emission will occur at the
endpoints to be offset by a similar one but opposite in sign from
that portion of the contour that continues to stretch out to infinity),
the final expressions for the amplitudes/cross-sections should be
multiplied by a factor of 1/2.
Equivalently, one might think of this compensation as actually
identifying the missing energy of the gluon emission at the
extremities with the off-mass-shellness. In fact, this is what we have
been implying all along when claiming that finite contours signify
off-mass-shellness.

Turning our attention to the contribution resulting from a virtual
gluon exchange from the entrance to the exit branch, with respect to
$z$, we consider the quantity
\begin{eqnarray}
  I_{3}
& = &
  \int_{-\sigma}^{0} dt_{1} \int_{0}^{\sigma} dt_{2}\,
  v^{\prime \mu} v^{\nu}
  D_{\mu\nu}\left( v^{\prime}t_{2}-vt_{1} \right)
=
  \frac{1}{4\pi^2}\left( -\pi\mu^{2} \right)^{(4-D)/2}
  \Gamma\left( {D\over 2}-1 \right)
\nonumber\\
&& \times
  v\cdot v^{\prime}\int_{0}^{\sigma} dt_{1}
  \int_{0}^{\sigma} dt_{2}
  \left(
          t_{1}^{2}v^{2} + t_{2}^{2} v^{\prime 2}
        + 2t_{1}t_{2} v\cdot v^{\prime} - i0_{+} \right)^{1-(D/2)}.
\label{eq:virtglex}
\end{eqnarray}
For case $(a)$ it assumes the form (recall that $v\cdot v^{\prime}$ is
negative for the DY and positive for the DIS type of process)
\begin{eqnarray}
  I_{3}^{(a)}
& = &
  \frac{1}{4\pi^{2}} \left( -\pi\mu^{2} \right)^{(4-D)/2}
  \Gamma\left( {D\over 2}-1 \right)
  \frac{v \cdot v^{\prime}}{|v||v^{\prime}|}
  \int_{0}^{1} dt_{1}
\nonumber \\
&&\times
   \int_{0}^{1} dt_{2}
  \left(  t_{1}^{2}+t_{2}^{2}+2t_{1}t_{2}
  \frac{v \cdot v^{\prime}}{|v||v^{\prime}|}
         -i0_{+} \right)^{1-(D/2)}.
\label{eq:virtglex2}
\end{eqnarray}

As shown in the Appendix, for DY, one determines ($\gamma_{\rm E}$ is
Euler's constant)
\begin{equation}
  I^{(a)}_{3,\text{DY}}
=
  \frac{1}{8\pi^{2}}\gamma_{\text{DY}}
  \coth\gamma_{\text{DY}}\frac{1}{2-{D\over 2}}
 +\frac{1}{8\pi^{2}}\gamma_{\text{DY}}
 \coth\gamma_{\text{DY}}
  \ln\left(\frac{\mu^{2}}{\lambda^{2}}
  \pi {\rm e}^{2+\gamma_{\text{E}}}\right),
\label{eq:IDY}
\end{equation}
where
$
 \cosh\gamma_{\text{DY}}
=
 w_{\text{DY}} = -\frac{v\cdot v^{\prime}}{|v||v^{\prime}|}\geq 1
$,
while for DIS
($w_{\text{DIS}}=\frac{v\cdot v^{\prime}}{|v||v^{\prime}|}\geq 1$)
\begin{equation}
  I^{(a)}_{3,{\rm DIS}}
=
   \frac{1}{8\pi^{2}}\gamma_{\text{DIS}}\coth\gamma_{\text{DIS}}
   \frac{1}{2-{D\over 2}}
  +\frac{1}{8\pi^{2}}\gamma_{\text{DIS}}\coth\gamma_{\text{DIS}}
   \ln\left(\frac{\mu^{2}}{\lambda^{2}}
   \pi {\rm e}^{2+\gamma_{\text{E}}}\right).
\label{eq:IDIS}
\end{equation}
In all of the above expressions, as well as in those that will
follow, we have ignored: (i) all imaginary terms that will drop out
when contributions (for virtual gluons) from the conjugate
contour are taken into account and (ii) finite, $\mu$-independent
terms that will cancel out when real-gluon contributions to the
cross-section are included.

Collecting all terms, we deduce, for the ``uniformly smooth'' part,
\begin{equation}
  I_{1}^{(a)}+I_{2}^{(a)}+I_{3}^{(a)}
=
    \frac{1}{8\pi^{2}}(\gamma \coth\gamma-1)
    \frac{1}{2-{D\over 2}}
  + \frac{1}{8\pi^{2}}(\gamma \coth\gamma-1)
    \ln\left(\frac{\mu^{2}}{\lambda^{2}}\pi {\rm e}^{2+\gamma_{\text{E}}}\right),
\label{eq:I123}
\end{equation}
where the angle $\gamma$ is to be adjusted either to the DY or the DIS
situation.

Concerning the ``jet'' part of the computation, we only need
to consider $I_{3}^{(b)}$ because \footnote{Recall the
remark following Eq.~(\ref{eq:I2}).} the expression for
$I_{1}^{(b)}+I_{2}^{(b)}$ is simply one half of that of
$I_{1}^{(a)}+I_{2}^{(a)}$.
A typical term entering
$I_{3}^{(b)}$ $(v_{-} \gg v^{\prime}_{+})$
is
\begin{equation}
  I_{3}^{(b)}
=
  \frac{1}{4\pi^{2}}\left( -\pi\mu^{2} \right)^{(4-D)/2}
  \Gamma\left( {D\over 2}-1 \right) v\cdot v^{\prime}
  \int_{0}^{\sigma} dt_{1} \int_{0}^{\sigma} dt_{2}
  \left( t_{1}^{2}v^{2}+2t_{1}t_{2}v\cdot v^{\prime}-i0_{+}
  \right)^{1-(D/2)},
\label{eq:I3}
\end{equation}
whose computation suffices to furnish each of the other three terms
as well.

It is shown in the latter part of the Appendix that for either the DY
or the DIS type process one obtains
\begin{equation}
  I_{3}^{(b)}
=
    \frac{1}{16\pi^{2}}\,\frac{1}{\left( 2-{D\over 2} \right)^{2}}
  + \frac{1}{16\pi^{2}}\,\frac{1}{2-{D\over 2}}
    \ln\left( \frac{\mu^{2}}{\lambda^{2}}\pi {\rm e}^{\gamma_{\rm E}}
       \right)
  + \frac{1}{32\pi^{2}}
    \ln ^{2}\left( \frac{\mu^{2}}{\lambda^{2}}
    \pi {\rm e}^{\gamma_{\text{E}}} \right)
  + \mbox{const}.
\label{eq:I3fin}
\end{equation}

Subtracting the pole terms in the $\overline{{\rm MS}}$ scheme, we
arrive at the finite part of the overall result.
For the uniformly soft contribution, in particular, we get
\begin{equation}
  (I_{1}^{(a)}+I_{2}^{(a)}+I_{3}^{(a)})_{\text{fin}}
=
  \frac{1}{8\pi^{2}}
\everymath{\displaystyle}
  \left( \gamma \coth\gamma-1 \right)
  \ln\left( \frac{\mu^{2}}{{\rule{0in}{2.0ex}\bar{\lambda}}^{2}}
     \right),
\label{eq:I123finsoft}
\end{equation}
while the jet contribution reads
\begin{equation}
  (I_{1}^{(b)}+I_{2}^{(b)}+I_{3}^{(b)})_{\text{fin}}
=
\everymath{\displaystyle}
  \frac{1}{16\pi^{2}}
  {\ln}^{2}\left( \frac{\mu^{2}}{{\rule{0in}{2.0ex}\bar{\lambda}}^{2}}
           \right),
\label{eq:I123finjet}
\end{equation}
where we have set
$\bar{\lambda}^{2}\equiv 4\lambda^{2} {\rm e}^{-2\gamma_{\rm E}}$.
The above relation takes into account all four different
configurations contributing to $I_{3}^{(b)}$.

Putting everything together we arrive at the following overall result for
the second order contribution stemming from contour $C^{z}$
\begin{equation}
\everymath{\displaystyle}
  U^{(2)}_{C,S}
=
  1 - \frac{\alpha_{\text{s}}}{2\pi} C_{\rm F}
  \left[ \gamma \, \coth \gamma \,
          \ln \left( \frac{\mu^{2}}
                          {{\rule{0in}{2.0ex}\bar{\lambda}}^{2}}
              \right)
        - {3\over 2} \ln \left( \frac{\mu^{2}}{{\rule{0in}{2.0ex}\bar{\lambda}}^{2}}
                         \right)
        + \ln ^{2} \left( \frac{\mu^{2}}{{\rule{0in}{2ex}\bar{\lambda}}^{2}}
                   \right)
  \right].
\label{eq:UfinCz}
\end{equation}
A similar result is obtained also for contour $\bar{C}^{z^{\prime}}$.

Noting that
$\gamma \coth \gamma = \ln \left( Q^{2}/m^{2} \right)$
(for $Q^{2}\gg m^{2}$),
with
$ Q^{2} = (p+p^{\prime})^{2} $
for the DY and
$Q^{2} = -(p^{\prime}-p)^{2}$
for the DIS type of process, respectively, we recognize that the
well known perturbative enhancements occurring as
$Q^{2} \rightarrow \infty$
are associated with the eikonal-type trajectories upon which our present
calculations have been based.
One, now, realizes that these trajectories define {\it threshold}
conditions, with respect to the given momentum exchange $Q$, for the
processes under consideration since they leave no room for space-time
contour fluctuations. In the following section we shall deal with the
summation of these enhanced contributions to leading logarithmic
order. We shall, furthermore, identify a correction factor associated
with those terms in Eq.~(\ref{eq:UfinCz}) not involving the
enhancement factor $\ln(Q^{2}/m^{2})$.

\section{Summation of enhanced contributions from virtual gluons}

The family of worldline paths to which the considerations in the
previous section have referred to recognize all (virtual) single-gluon
exchanges, consistent with the simple geometrical configuration of
two constant four-velocities making a fixed angle $\gamma$ between them
(in Euclidean formulation). Among these gluons there will be ``hard''
ones (upper limit $Q$) and ``soft'' ones (lower limit set by
$\bar{\lambda})$.
What is debited to the former and what to the latter group of gluons
is, of course, relative. It is precisely the role of the
renormalization scale $\mu$, entering through the need to face UV
divergences arising even for the restricted family of paths, to
provide the dividing line. The corresponding renormalization-group
equation reflects the fact that the scale $\mu$ is arbitrary and that
physical results do not depend on it. A straightforward application of
this fact will enable us to sum the enhanced, virtual gluon
contribution to the amplitude in leading logarithmic order as well as
to obtain a bona-fide correction term.

To bring the above discussion into a concrete form, consider a
factorization, good to order $1/Q^{2}$, between the ``soft'' and the
``hard'', relative to the scale $\mu$, gluon contributions to the
amplitude $U_{C}$ entering Eq.~(\ref{eq:wilson}) for the restricted set
of contours.\footnote{For the purposes of the following discussion, it
suffices to use the generic notation $C$ for either contour.}
We write
\begin {equation}
\everymath{\displaystyle}
  U_{C}
=
  U_{C,S}\left ({Q^{2}\over m^{2}},{\mu^{2}\over {\rule{0in}{2ex}\bar{\lambda}}^{2}}
         \right)
  U_{C,H}\left( {Q^{2}\over \mu^{2}} \right)+{\cal O}\left( {1\over Q^{2}} \right)
\label{eq:Ufact}
\end{equation}
noting that the soft part $U_{C,S}$ coincides, to second order, with
$U_{C,S}^{(2)}$.

It is further convenient to separate the part of the ``soft'' term
involving the enhancement factor from the one which does not.
To this end we write, to second order in perturbation theory,
\begin{equation}
  U_{C,S}^{(2}
=
  U_{C,\text{cusp}}^{(2)}U_{C,\text{coll}}^{(2)}
\label{eq:Usep}
\end{equation}
with
\begin{equation}
\everymath{\displaystyle}
  U_{C,\text{cusp}}^{(2)}
=
  1-\frac{\alpha_{\text{s}}}{2\pi}C_{\rm F}
  \ln\left( {Q^{2}\over m^{2}} \right)
  \ln\left( {\mu^{2}\over {\rule{0in}{2ex}\bar{\lambda}}^{2}} \right)
\label{eq:Ucusp}
\end{equation}
and
\begin{equation}
\everymath{\displaystyle}
  U_{C,\text{coll}}^{(2)}
=
  1-\frac{\alpha_{\text{s}}}{2\pi}C_{\rm F}
  \left[ \ln ^{2}
  \left(
  {\mu^{2}\over{\rule{0in}{2ex}\bar{\lambda}}^{2}}
  \right)
 -{3\over 2}\ln\left( {\mu^{2}\over{\rule{0in}{2ex}\bar{\lambda}}^{2}}
        \right)
 \right].
\label{eq:Ucoll}
\end{equation}
In the above relations, the designation ``cusp'' refers to that part
of the soft factor which recognizes the angle $\gamma$, whereas the
characterization ``collinear'' signifies independence from $\gamma$.
From a physical standpoint, it represents the contribution from those
gluons which are emitted collinearly with incoming or outgoing
four-velocities due to the sudden acceleration/deceleration occurring
at the cusp point. It should be further pointed out that collinear
emission occurring in the above context pertains to the soft sector
and does not coincide in the full with what is designated as
``collinear'' by Collins et al. in \cite{CSS89}.
Full agreement on ``collinearity'' will arise a posteriori, i.e., once
resummation is performed.

From Eqs.~(\ref{eq:Ufact})-(\ref{eq:Ucoll}) we obtain a relation which
does not involve $U_{\rm coll}$ by simply taking the logarithmic
derivative with respect to $Q^{2}$:
\begin{equation}
  \frac{d}{d\ln Q^{2}}\ln U
=
  \frac{d}{d\ln Q^{2}}\ln U_{\text{cusp}}+\frac{d}{d\ln Q^{2}}\ln U_{\rm H},
\label{eq:noUcoll}
\end{equation}
where we have dropped the subscript $C$ as being superfluous in the
considerations to follow.

The $\mu$-independence of physical results leads to the
renormalization group equation
\begin{equation}
  \frac{d}{d\ln\mu}\frac{d}{d\ln Q^{2}}\ln U_{\rm H}
=
  -\frac{d}{d\ln\mu}\frac{d}{d\ln Q^{2}}
   \ln U_{\text{cusp}}
=
  -{1\over 2}\Gamma_{\text{cusp}}(\alpha_{\text{s}})
\label{eq:muindep}
\end{equation}
with $\Gamma_{{\rm cusp}}$ to be read off from
Eqs.~(\ref{eq:I123})-(\ref{eq:I3fin}) and (\ref{eq:U2order});
specifically,
\begin{equation}
  \Gamma_{\text{cusp}}(\alpha_{\text{s}})
=
  {\alpha_{\text{s}}\over\pi}C_{\rm F}+{\cal O}(\alpha_{\text{s}}^{2}).
\label{eq:Gammacusp}
\end{equation}

From the second leg of Eq.~(\ref{eq:muindep}), one obtains
\begin{equation}
  \frac{d}{d\ln Q^{2}}\ln U_{\text{cusp}}
=
  -\int_{\bar{\lambda}^{2}}^{\mu^{2}}\frac{dt}{2t}
   \Gamma_{\text{cusp}}\left[ \alpha_{\text{s}}(t) \right]
\label{eq:cuspcusp}
\end{equation}
which, in turn, gives
\begin{equation}
  \frac{d}{d\ln Q^{2}}\ln U_{\rm H}
=
  -\int_{\mu^2}^{Q^2}\frac{dt}{2t}
  \Gamma_{\text{cusp}}\left[ \alpha_{\text{s}}(t) \right]+
  \Gamma\left[ \alpha_{\text{s}}(Q^{2}) \right],
\label{eq:cuspH}
\end{equation}
where we have defined
\begin{equation}
  \Gamma\left[ \alpha_{\text{s}}(Q^{2}) \right]
\equiv
  \frac{d}{d\ln Q^{2}}\ln U_{\rm H}
  \left( {Q^{2}\over\mu^{2}}
  \right)_{\mu^{2}=Q^{2}}.
\label{eq:defGamma}
\end{equation}

Combining the last three equations we have
\begin{equation}
  \frac{d}{d\ln Q^{2}}\ln U
=
  -\int_{\bar{\lambda}^{2}}^{Q^{2}}\frac{dt}{2t}
    \Gamma_{\text{cusp}}\left[ \alpha_{\text{s}}(t) \right]
   +\Gamma\left[ \alpha_{\text{s}}(Q^{2}) \right],
\label{eq:Gammatot}
\end{equation}
where, in terms of $\ln U$, we write
\begin{equation}
  \Gamma\left[ \alpha_{\text{s}}(Q^{2}) \right]
\equiv
  \frac{d}{d\ln Q^{2}}\ln U|_{\bar{\lambda}^{2}=Q^{2}}.
\label{eq:Gamma}
\end{equation}

Setting $Q^{2}=\mu^{2}$, we are led to the isolation of the ``cusp''
and ``collinear'' terms according to
\begin{equation}
\everymath{\displaystyle}
  \ln U
=
   \ln U_{\text{cusp}}\left(
                           {Q^{2}\over m^{2}},{Q^{2}
                           \over {\rule{0in}{2ex}\bar{\lambda}}^{2}}
                     \right)
 + \ln U_{\text{coll}}
 \left(
       {Q^{2}\over {\rule{0in}{2ex}\bar{\lambda}}^{2}}
 \right)
 + \ln U_{\rm H}\left( \frac{Q^{2}}{\mu^{2}}\right)_{\mu^{2}=Q^{2}}
\label{eq:cusppcoll}
\end{equation}
from which, once appealing to Eq.~(\ref{eq:cuspcusp}), we obtain
\begin{equation}
  \frac{d}{d\ln Q^{2}}\ln U|_{\bar{\lambda}^{2}=Q^{2}}
=
  \frac{d}{d\ln Q^{2}}\ln U_{\text{coll}}|_{\bar{\lambda}^{2}=Q^{2}}
\label{eq:UUcoll}
\end{equation}
which leads to the identification
\begin{equation}
  \Gamma\left[ \alpha_{\text{s}}(Q^{2}) \right]
=
  \frac{d}{d\ln Q^{2}}\ln U_{\text{coll}}|_{\bar{\lambda}^{2}=Q^{2}}
=
  {3\over 4}\alpha_{\text{s}}(Q^{2})+{\cal O}(\alpha_{\text{s}}^{2}).
\label{eq:GammaUcoll}
\end{equation}

Collecting together all our findings, we obtain our final, re-summed
result corresponding to the contour $C$.
It reads
\begin{equation}
  U
=
  \exp\left\{-\int_{\bar{\lambda}^{2}}^{Q^{2}}\frac{dt}{2t}
       \left[ \ln{Q^{2}\over t}\Gamma_{\text{cusp}}(\alpha_{\text{s}}(t))
      -\Gamma(\alpha_{\text{s}}(t))
      \right]
      \right\}
  U_{0}(\alpha_{\text{s}}(Q^{2})).
\label{eq:resU}
\end{equation}

One notes that the correction factor is associated with collinear
emission (cf. Eq.~(\ref{eq:GammaUcoll})). Secondly, the factor
$U_{0}(\alpha_{\text{s}}(Q^{2}))$
represents initial conditions input {\it at the QCD level}.
One cannot help but bring to mind evolution equations at the partonic
level where the probabilistic interpretation prevails.
The connection between evolution and renormalization group equations
is widely taken for granted.
The worldline approach might provide the space-time description
framework for understanding the precise connection between the two
types of equations.
Finally, let us remark that the conjugate-contour term
$U^{\dagger}(\bar{C}^{z^{\prime}})$
can be treated in a completely analogous fashion.

\section{Summation of enhanced contributions from real gluon emission}

\vspace{.5cm}

\begin{figure}
\centering \epsfig{file=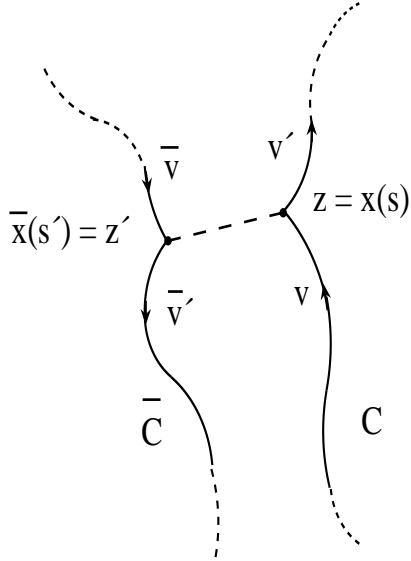,height=8cm,width=6.0cm,silent=}
\vspace{0.2cm}
\caption{\footnotesize
                Neighborhoods of respective points on two
                conjugate contours, where the momentum transfer
                takes place, and associated four-velocities.}
\label{fig:paths3}
\end{figure}

We shall now turn our attention to real gluons and attempt to
factorize cross-section contributions from neighborhoods around
points $z$ and $z^{\prime}$.
Note that we now have to deal with gluons which connect two
``opposite'' neighborhoods while crossing the unitarity line.
(this situation is depicted in Fig.~\ref{fig:paths3})
The relevant scale promptly entering our considerations is the impact
parameter $b=z-z^{\prime}$, which must be eventually integrated over
in order to get the cross-section. Naturally, the short-distance
cutoff in this integration will be provided by the (length) scale
$1/|Q|$.

%

For the eikonal-type family of paths and in first order perturbation
theory, the relevant quantity on which our quantitative considerations
are to be based, i.e., the counterpart of Eq.~(\ref{eq:U2order}),
is given by
\begin{eqnarray}
  U^{(2)}_{C\bar{C},S}
=
  1
& + &
  g^{2}C_{\rm F}\left[
                      \int_{-\sigma}^{0} dt_{1}
                      \int_{-\sigma}^{0} dt_{2}
                      v\cdot\bar{v}\, D\left( t_{1}v-t_{2}\bar{v}+b
                                       \right)\right.
                      \nonumber \\
& + & \left.
                      \int_{0}^{\sigma} dt_{1}
                      \int_{0}^{\sigma} dt_{2}
                      v^{\prime}\cdot\bar{v}^{\prime}\,
                      D\left( t_{1}v^{\prime}-t_{2}\bar{v}^{\prime}+b
                      \right)\right.
\nonumber\\
& + & \left.
                      \int_{-\sigma}^{0} dt_{1}
                      \int_{0}^{\sigma} dt_{2} v\cdot\bar{v}^{\prime}\,
                      D\left( t_{1}v-t_{2}\bar{v}^{\prime}+b
                      \right)\right.
\nonumber \\
& + & \left.
                      \int_{0}^{\sigma} dt_{1}
                      \int_{-\sigma}^{0} dt_{2} v^{\prime}\cdot\bar{v}\,
                      D\left( t_{1}v^{\prime}-t_{2}\bar{v}+b
                      \right)
               \right],
\label{eq:U2orderreal}
\end{eqnarray}
where the bar denotes four-velocities for the conjugate contour and
the subscript cut is henceforth omitted.

To identify the leading behavior of $U^{(2)}_{C\bar{C},S}$, with
respect to $b$, we shall consider first the situation corresponding
to $b=0$.
The subsequent emergence of UV divergences, once handled through
dimensional regularization, will introduce a mass scale $\bar{\mu}$
that will be bounded from below by an IR cutoff $\lambda$ and from
above by the (mass) scale $1/b$. The resulting renormalization group
equation will facilitate the resummation of the leading terms, just
as in the virtual-gluon case.

Let us commence our quantitative considerations by looking at the term
\begin{equation}
  J_{1}(b)
\equiv
  v\cdot\bar{v} \int_{-\sigma}^{0} dt_{1}
  \int_{-\sigma}^{0} dt_{2}\, D\left( t_{1}v-t_{2}\bar{v}+b \right)
\label{eq:Jb}
\end{equation}
with $v^2=\bar{v}^2= -v\cdot\bar{v}$ (see Fig.~\ref{fig:paths3}).

Setting $b=0$ and using the expression for the cut propagator as given
by Eq.~(\ref{eq:realgluonemission}), we obtain
\begin{equation}
  J_{1}(0)
=
  -{1\over 4\pi^{2}}
  \left( -\pi\bar{\mu}^{2}L_{1}^{2} \right)^{\left(2-D/2\right)}
  \Gamma\left({D\over 2}-1\right)\,{1\over D-3}\,{1\over 4-D}
  \left[ 1-(2^{4-D}-1) \right]
\label{eq:J0}
\end{equation}
which actually coincides with what one would obtain if the regular
propagator was substituted. The significance of this occurrence is
that it leads to the same anomalous dimensions for the running of
the real gluon contribution to the cross-section as for the virtual
part. This fact can be immediately verified via a direct comparison
with Eq.~(\ref{eq:I1}).

Isolating the finite part of the above expression, we write
\begin{equation}
  J_{1}^{(a),\text{fin}}
=
  -{1\over 8\pi^{2}} \ln\left({\bar{\mu}^{2}\over\lambda^{2}}\right).
\label{eq:J1fin}
\end{equation}
It is trivial to see that the same result holds also for
$J_{2}^{(a),{\rm fin}}$.

We next turn our attention to the term
\begin{equation}
  J_{3}(b)
\equiv
     v\cdot\bar{v}^{\prime} \int_{-\sigma}^{0} dt_{1}
     \int_{0}^{\sigma} dt_{2}\,D\left( t_{1}v-t_{2}\bar{v}^{\prime}
   + b \right).
\label{eq:J3b}
\end{equation}
Its computation will concurrently allow us to determine $J_{4}(b)$
which corresponds to the exchange prime$\leftrightarrow$ no-prime in
the expression above.

Dimensionally regularizing the cut propagator, we obtain
\begin{eqnarray}
  J_{3}(0)
& = &
  {1\over 4\pi^{2}}(-\pi\bar{\mu}^{2})^{(4-D)/2}
  \Gamma\left( {D\over 2}-1 \right)
  v\cdot\bar{v}^{\prime}\!\!\int_{0}^{\sigma} dt_{1}
\nonumber \\
&& \times
  \int_{0}^{\sigma}\! dt_{2}
  \left( t_{1}^{2}v^{2}+t_{2}^{2}\bar{v}^{\prime
  2}+2v\cdot\bar{v}^{\prime}t_{1}t_{2}
  -i0_{+} \right)^{1-D/2}.
\label{eq:cutprop}
\end{eqnarray}
Once again we record, by referring to Eq.~(\ref{eq:I3}), coincidence
of the singularities and, by extension, of associated anomalous
dimensions between virtual and real gluon expressions that contribute
to the cross section.

For the ``uniformly soft'' configuration the corresponding result is
\begin{eqnarray}
  J_{3}^{(a)}(0)
& = &
  {1\over 4\pi^{2}}(-\pi\bar{\mu}^{2})^{(4-D)/2}
  \Gamma\left( {D\over 2}-1 \right)\!
  \frac{v\cdot\bar{v}^{\prime}}{|v||\bar{v}^{\prime}|}
\nonumber \\
&& \times
  \int_{0}^{1}\! dt_{1}\!
  \int_{0}^{1}\! dt_{2}\!
  \left( t_{1}^{2}+t_{2}^{2}+2t_{1}t_{2}
  \frac{v\cdot\bar{v}^{\prime}}{|v||\bar{v}^{\prime}|}-i0_{+}
  \right)^{1-D/2}.
\label{eq:Jsoft}
\end{eqnarray}

Taking into consideration that for the DY case
$
 \frac{v\cdot\bar{v}^{\prime}}{|v||\bar{v}^{\prime}|}
=
 \frac{v^{\prime}\cdot\bar{v}}{|v^{\prime}||\bar{v}|}
=
 \cosh\gamma_{\rm DY} > 0
$
we obtain
\begin{eqnarray}
  J_{3,\text{DY}}^{(a)}(0)
=
  J_{4,\text{DY}}^{(a)}(0)
& = &
  {1\over 4\pi^{2}}(-\pi\bar{\mu}^{2})^{(4-D)/2}
  \Gamma\left( {D\over 2}-1 \right)
  \cosh\gamma_{\text{DY}}
\nonumber\\
&& \times
  \int_{0}^{1} dt_{1}
  \int_{0}^{1} dt_{2} \,
  \left( t_{1}^{2}+t_{2}^{2}+2t_{1}t_{2}
  \cosh\gamma_{\text{DY}}-i0_{+} \right)^{1-D/2},
\label{eq:J3DY}
\end{eqnarray}
whose finite part reads
\begin{equation}
  J_{3,\text{DY}}^{(a),\text{fin}}(0)
=
  J_{4,\text{DY}}^{(a),\text{fin}}(0)
=
  {1\over 8\pi^{2}}
  \gamma_{\text{DY}}\coth\gamma_{\text{DY}}\,
  \ln\left( {\bar{\mu}^{2}\over\lambda^{2}} \right).
\label{eq:J3DYfin}
\end{equation}

For the DIS case, where
$
 \frac{v\cdot\bar{v}^{\prime}}{|v||\bar{v}^{\prime}|}
=
 \cosh(\gamma_{\text{DIS}}-i\pi)<0
$,
and
$
 \frac{v^{\prime}\cdot\bar{v}}{|v^{\prime}||\bar{v}|}
=
 \cosh(\gamma_{\text{DIS}}+i\pi)<0$,
we find
\begin{equation}
  J_{3,\text{DIS}}^{(a),\text{fin}}(0)
=
  {1\over 8\pi^{2}}\left( \gamma_{\text{DIS}}-i\pi \right)
  \coth\gamma_{\text{DIS}}\,\ln\left( {\bar{\mu}^{2}\over\lambda^{2}}
                              \right),
\label{eq:J3DIS}
\end{equation}
whereas
\begin{equation}
  J_{4,\text{DIS}}^{(a),\text{fin}}(0)
=
  {1\over 8\pi^{2}}\left( \gamma_{\text{DIS}}+i\pi \right)
  \coth\gamma_{\text{DIS}}\,\ln\left( {\bar{\mu}^{2}\over\lambda^{2}} \right).
\label{eq:J4DIS}
\end{equation}

Given that the imaginary parts in the last two equations will
eventually drop out, from now on we shall write, generically,
\begin{equation}
  J_{3}^{(a),\text{fin}}(0)
=
  J_{4}^{(a),\text{fin}}(0)
=
 {1\over 4\pi^{2}}
 \gamma \coth\gamma\,\ln\left( {\bar{\mu}^{2}\over\lambda^{2}}
 \right),
\label{eq:J3=J4}
\end{equation}
where $\gamma$ can be adjusted to the appropriate case.

Turning our attention to the ``jet'' configuration, we can
go directly to
$J_{3}^{(b)}(0)$,
since
$J_{1}^{(b)}(0)+J_{2}^{(b)}(0)$
furnishes half the contribution of its uniformly soft counterpart,
the reason being the same as the one given in the virtual gluon case.
We thus have
\begin{equation}
  J_{3}^{(b)}(0)
=
  -4{1\over 4\pi^{2}}\left( -\pi\bar{\mu}^{2} \right)^{(4-D)/2}
  \Gamma\left( {D\over 2}-1 \right)
  \frac{v\cdot\bar{v}^{\prime}}{|v|}
  \int_{0}^{1} dt_{1}
  \int_{0}^{1} dt_2\, \left(t_{1}^{2}+2t_{1}t_{2}
                            \frac{v\cdot\bar{v}^{\prime}}{|v|}-i0_{+}
                      \right)^{1-D/2}
\label{eq:J3jet}
\end{equation}
with a corresponding expression holding also for
$J_{4}^{(b)}(0)$.

For the finite parts one obtains
\begin{equation}
  J_{3}^{(b),\text{fin}}(0)+J_{4}^{(b),\text{fin}}(0)
=
  {1\over 4\pi^{2}} \ln ^{2}
  \left( {\bar{\mu}^{2}\over\lambda^{2}} \right),
\label{eq:J3+4fin}
\end{equation}
which holds true for both the DY and the DIS case.

Collecting our findings from the real-gluon analysis to the
second-order level, we write for the finite contribution to the
cross-section
\begin{eqnarray}
  U_{C\bar{C},S^{(2)}}
& = &
  1
+
  \frac{\alpha_{\text{s}}}{\pi}C_{\rm F}
  \left[
         \gamma \coth\gamma\,\ln
         \left( {\bar{\mu}^{2}\over\lambda^{2}}
         \right)
        -{3\over 2}\ln\left( {\bar{\mu}^{2}\over\lambda^{2}} \right)
        +\ln ^{2}\left( {\bar{\mu}^{2}\over\lambda^{2}} \right)
  \right]
\nonumber\\
& = &
  U_{C\bar{C},\text{cusp}}^{(2)}U_{C\bar{C},\text{coll}}^{(2)} \; ,
\label{eq:UcuspUcoll}
\end{eqnarray}
where (to the order we have been calculating) the results for the
``cusp'' and the ``collinear'' terms read
\begin{equation}
  U_{C\bar{C},\text{cusp}}
=
  1+\frac{\alpha_{\text{s}}}{\pi}C_{\rm F}
  \gamma \coth\gamma\, \ln\left( \bar{\mu}^{2}\over\lambda^{2} \right)
\label{eq:Ucuspfin}
\end{equation}
and
\begin{equation}
  U_{C\bar{C},\text{coll}}^{(2)}
=
  1+\frac{\alpha_{\text{s}}}{\pi}C_{\rm F}
  \left[
        -{3\over 2}\ln\left( {\bar{\mu}^{2}\over\lambda^{2}} \right)
        +\ln ^{2}\left( {\bar{\mu}^{2}\over\lambda^{2}} \right)
  \right].
\label{eq:Ucollfin}
\end{equation}

At the same time, the singularity structure of the full expression
for the cross-section entails a multiplicative renormalization factor,
which is common to all ``Wilson loop'' configurations entering its
description, but which is the {\it only} one that pertains to the
family of eikonal-type paths under consideration.
The reasoning is, of course, identical to the one promoted for the
virtual gluon case. Therefore, the corresponding contribution
to the cross-section factorizes and the same resummation procedure
can be employed as for the virtual-gluon case.
As already observed, the anomalous dimension is in both cases the
same.
There are, however, the following notable differences.
First, the upper limit for the momentum of real-gluon emission is
$1/b^{2}$ instead of $Q^{2}$.
Second, there is a difference of sign, which becomes evident
by comparing Eqs.~(\ref{eq:Ucusp}) and (\ref{eq:Ucoll}) with
Eqs.~(\ref{eq:Ucuspfin}) and (\ref{eq:Ucollfin}) and, finally, no hard
real-gluon emission enters our considerations - by definition.
In this light, it is practically self-evident that the resummed
expression for real-gluon emission becomes
\begin{equation}
  U_{C\bar{C}}
=
  \exp\left\{ \int_{\bar{\lambda}^{2}}^{1/b^{2}}
  \frac{dt}{t}\left[ \ln{Q^{2}\over t}
                      \Gamma_{\text{cusp}}\left( \alpha_{\text{s}}(t) \right)
                     -\Gamma\left( \alpha_{\text{s}}(t) \right)
              \right]
      \right\}
              U_{C\bar{C},0}.
\label{eq:res_real_gl_emi}
\end{equation}

We can now bring together real and virtual gluon results by referring
to our generic expression for the cross-section as given by
Eq.~(\ref{eq:wilson}).
We write
\begin{equation}
  {\cal W}
=
  \left\langle Tr \left( U^{\dagger} ( \bar{C}^{z^{\prime}})
  U(C^z) \right)
  \right\rangle
=
  U_{C}U_{\bar{C}}U_{C\bar{C}},
\label{eq:3Us}
\end{equation}
where we have first extracted the resumed expression corresponding
to the curly bracket expectation values from virtual gluon
exchanges in the amplitude and its conjugate. Clearly, the last
factor corresponds to the resummed expression resulting from
the real-gluon ``averaging''.

Now, at the cross-section level, our threshold resummation
of the virtual gluons reads
\begin{equation}
  U_{C}U_{\bar{C}}
=
  \exp\left\{-\int_{\bar{\lambda}^{2}}^{Q^{2}}
  \frac{dt}{t}
  \left[ \ln{Q^{2}\over t}\Gamma_{\text{cusp}}(\alpha_{\text{s}}(t))
        -\Gamma(\alpha_{\text{s}}(t))
  \right]
      \right\}
              U_{C,0}U_{\bar{C},0}.
\label{eq:rescrossecvirt}
\end{equation}

Finally, by making the re-adjustment
$
 \ln\left( \bar{\mu}^{2}/\lambda^{2} \right)
\rightarrow
 \ln\left( \mu^{2}/\bar{\lambda}^{2} \right)
$ we obtain our final result, which reads
\begin{equation}
  {\cal W}_{C\bar{C}}
=
  \exp\left\{ -\int_{c/b^{2}}^{Q^{2}}\frac{dt}{t}
  \left[
         \ln{Q^{2}\over t} \Gamma_{{\rm cusp}}(\alpha_{\text{s}}(t))
        -\Gamma(\alpha_{\text{s}}(t))
  \right]
      \right\}{\cal W}_{0},
\label{eq:finres}
\end{equation}
with $\Gamma_{{\rm cusp}}$ and $\Gamma$ given by
Eqs.~(\ref{eq:Gammacusp}) and (\ref{eq:GammaUcoll}),
respectively and where
$c=4{\rm e}^{-2\gamma_{\rm E}}$.
an expression whose previous derivation, see, e.g., Ref.~\cite{KS95}
and references therein, has employed Wilson lines, as an exogenous
element attached to quark operators

\section{Concluding remarks}

In this paper we have studied the threshold resummation behavior of
DY and DIS type of processes, widely pursued by Sterman and others
(see, for example, in \cite{Ster87,CLS97}), by staying strictly
within the framework of QCD.
From a physical viewpoint, we based our considerations on
soft-gluon radiation, restricting ourselves to an energy regime
whose lower cutoff is high enough to justify an analysis in which
reference to ``gluons'', as dynamical degrees of freedom,
continues to make sense.
In this context we have implicitly assumed the pre-confinement
property, originally articulated in the first work of
reference \cite{AV79} (see also \cite{BCM80}), according to which
the non-perturbative dynamics responsible for confinement screens
color up to the infrared scale $\lambda$ which sets the lower
limit for the perturbative regime.

Our analysis has been conducted via the adoption of the worldline
casting of QCD whose emphasis on a space-time propagation content
facilitates the isolation and eventually factorization of IR
contributions to cross sections through eikonal-type paths.
In this connection, let us remark that we have already extensively
applied this approach to the calculation of various fermionic Green's
functions \cite{KKS92} (see, also \cite{KS89}), the Sudakov form
factor \cite{GKKS97}, as well as to derive gluon Reggeization in
connection with the four-point fermionic Green's function - for the latter
case in the forward regime \cite{KK98}.
The present paper presents the first application of the worldline
methodology to ``cross-sections'' in which real-gluon radiation effects
must also be taken into account - and so far have received much less
systematic treatment.

The common renormalization-group running of virtual and real
- uniformly soft and jet-type - gluon contributions to the cross-section
has led to a final result in which the IR-cutoff momentum is provided
by the inverse impact parameter which can be vested with concrete physical
meaning related to the process itself. A number of possible physical
applications, ranging from extensions to exclusive processes to
intrusions into the non-perturbative regime of QCD, as per, for example,
along the lines of Ref.~\cite{KS95} come to mind which will occupy our
attention in the immediate future.

\newpage
\appendix
\section{}

Our task is to establish Eqs.~(\ref{eq:IDY}) and (\ref{eq:IDIS}) in
the text. Performing the integration entering the right hand side of
Eq.~(\ref{eq:virtglex2}), one obtains
\begin{eqnarray}
  I_{3,\text{DY}}^{(a)}
&=&
  \frac{1}{4\pi^{2}}\left( -\pi\mu^{2} \right)^{(4-D)/2}
  \Gamma\left( {D\over 2}-1 \right)
  {1\over 4-D}\, {1\over D-3}\, 2w_{\text{DY}}
\nonumber \\
&& \times
  \left\{ w_{\text{DY}}
  F\left( 1,{D\over 2}-1;{D-1\over 2};1-w_{\text{DY}}^{2} \right)
  \right.
\nonumber \\
&&
\left.
  +{1\over 2}\left[ 2(1-w_{\text{DY}}) \right]^{2-D/2}
   F\left( 1,{D\over 2}-1;{D-1\over 2};{1+w_{\text{DY}}\over 2} \right)
 \right\}
\label{eq:A.1}
\end{eqnarray}
for DY kinematics and
\begin{eqnarray}
  I_{3,\text{DIS}}^{(a)}
&=&
  \frac{1}{4\pi^{2}}\left( -\pi\mu^{2} \right)^{(4-D)/2}
  \Gamma\left( {D\over 2}-1 \right)
  {1\over 4-D}\, {1\over D-3}\, 2w_{\text{DIS}}
\nonumber \\
&& \times
  \left\{ w_{\text{DIS}}
  F\left( 1,{D\over 2}-1;{D-1\over 2};1-w_{\text{DIS}}^{2} \right)
  \right.
\nonumber\\
&&
\left.
  -{1\over 2}\left[ 2(1+w_{\text{DIS}})\right]^{2-D/2}
   F\left( 1,{D\over 2}-1;{D-1\over 2};{1-w_{\text{DIS}}\over 2} \right)
  \right\}
\end{eqnarray}
for the DIS situation.

Setting $D=4$, we obtain
\begin{equation}
  F\left( 1,1;3/2;1-w_{\text{DY}}^{2} \right)
=
  \frac{\gamma_{\text{DY}}}{\sinh\gamma_{\text{DY}}\cosh\gamma_{\text{DY}}}
\end{equation}
and
\begin{equation}
   F\left( 1,1;3/2;{1+w_{\text{DY}}\over 2} \right)
=
   \frac{\gamma_{\text{DY}}}{\sinh\gamma_{\text{DY}}}
  -i\frac{\pi}{\sinh\gamma_{\text{DY}}}.
\end{equation}
As the imaginary part in the above expression will cancel against its
counterpart in the conjugate expression, it can be dropped as far as the
cross-section is concerned.

Denoting the expression inside the curly brackets on the rhs
of Eq.~(\ref{eq:A.1}) by $f_{D}^{\text{DY}}(w)$ and setting
\begin{equation}
    f_{D}^{\text{DY}}(w)
=
    f_{4}^{\text{DY}}(w)+(4-D)\frac{f_{D}^{\text{DY}}(w)
  - f_{4}^{\text{DY}}(w)}{4-D},
\end{equation}
we realize that the second term on the rhs will lead to finite terms
that depend solely on $w$ and which will cancel from similar
contributions of the same sort coming from the other terms entering
Eq.~(\ref{eq:U2order}).
Putting everything together, one finally arrives at
Eq.~(\ref{eq:IDY}).
Clearly, the DIS situation is similarly confronted.

To establish the result given by Eq.~(\ref{eq:I3fin}), we first
note that Eq.~(\ref{eq:I3}) gives
\begin{eqnarray}
  I_{3}^{(b)}
&=&
  \frac{1}{4\pi^{2}}
  \left( -\pi{\mu^{2}\over\lambda^{2}} \right)^{(4-D)/2}
  \Gamma\left( {D\over 2}-1 \right)
  {1\over (4-D)^{2}}
  \left( \frac{2v\cdot v^{\prime}}{|v|^{2}} \right)^{(4-D)/2}
\nonumber\\
&& \times
   \left[
         F\left( {D\over 2}-1,2-{D\over 2};
                3-{D\over 2};-{2v\cdot v^{\prime}\over |v|^2}
          \right)
       + \left( 1+{2v\cdot v^{\prime}\over |v|^{2}}
       \right)^{2-D/2}\right.
\nonumber \\
&&
\left.
      \quad\;\; - \left( {2v\cdot v^{\prime}\over |v|^{2}} \right)^{2-D/2}
   \right].
\end{eqnarray}

Now, the distinction between the DY and DIS like processes resides in
the sign of $v\cdot v^{\prime}$ (positive and negative, respectively).
On the other hand, what {\it is} relevant for the ``jet'' situation
is that the absolute value of $2{v\cdot v^{\prime}\over |v|^{2}}$ is of the
order of unity. One, then, easily determines that in the limit
$D\rightarrow 4$ Eq.~(\ref{eq:I3fin}) is retrieved.

\end{document}